# A Sign in Space Part 2, a new simulated First Contact scenario engaging the global scientific SETI network and integrating deep-space technologies into SETI search.

Daniela de Paulis (SETI Institute)
Andrea Melis (INAF)
Daniel Estévez (SETI Institute)

## Abstract

The Search for Extraterrestrial Intelligence programme aims to find evidence of extraterrestrial intelligence. SETI scientists pursue this objective by searching for technosignatures primarily in the radio or optical spectrum. The IAA SETI Committee has been developing protocols to offer guidelines in case of an actual extraterrestrial signal reception. These protocols establish procedures to confirm the authenticity of the signal and its actual extraterrestrial nature, as well as the committee's role in managing the discovery in relation to the scientific community and the general public, targeting scientific queries in relation to the management and development of equipment and methodologies employed in current SETI search. The project A Sign in Space part 2 aims to actively engage the global SETI community and the global telescope network, with scientific facilities responding and coordinating themselves to confirm the extraterrestrial nature of the received signal. The project develops some of the societal, scientific and technological methodologies already employed in the interdisciplinary work A Sign in Space. Here, we established a global collaboration between SETI scientists, radio observatories and the European Space Agency to test the communication and technological pipeline and protocols leading to the reception of a simulated extraterrestrial signal by a network of radio telescopes across the globe. We also employed an ESA spacecraft for system validation. A Sign in Space part 2 aims to further implement deep space technologies in SETI search through a structured and sustained programme, in collaboration with ESA, the SETI Institute and the Italian Institute for Astrophysics. The project will aim to stage and test the entire Post Detection procedure, implementing the protocol as envisioned by the IAA SETI committee.

## Project's Overview

A Sign in Space part 2 aims to support the global SETI scientific community by providing a testing ground for current signal detection methodologies. The outcomes of these tests will generate valuable feedback that the SETI scientific community can use to improve current SETI scientific tools and procedures.
We are aiming to tackle the following:

- Test the prompt and effective technical reception capabilities of a SETI signal by a global radio antenna network. Testing strengths and weaknesses of the current system will in turn help create a more robust global network, optimising prompt signal detection, signal processing and communication of a SETI candidate signal across the network;
- Help bridge the gap between signal detection and organisational response, supporting the international community in reacting effectively to a potential extraterrestrial discovery;
- Apply and test as accurately as possible the Post Detection Protocol designed by the International Academy of Astronautics SETI Committee in 2025-2026 to identify strengths and areas for improvement;
- Support the design of a global Detection Protocol to ensure possible candidate signals are optimally received by the global telescope network;
- Help expand the current SETI telescope network to a worldwide scale;
- The project is an excellent opportunity to deploy deep-space technologies for improving SETI search from Earth and might help establish a new direction in the scientific field;
- With this project, we focus on one of the several search methodologies pursued by SETI, which is looking for a possible radio signal from an extraterrestrial civilisation that shares similar technological knowledge as humans. We will support the SETI scientific community in testing the limits of current procedures and technology to help strengthen the global detection pipeline and optimise the equipment designated for SETI search.

**Project's Key Objectives**

- Testing the pipeline: how to detect a potential ET signal and how to activate the global network pipeline of radio telescopes receiving at different frequencies.

  Current radio astronomy and SETI facilities mostly analyse data in post-processing rather than in real time, often several months after the observations. There is a gap both in technology and community organisation between the ideal scenario of real-time detections and low-latency triggering of other facilities, which would be required for a SETI detection and the current situation. Also, global communication issues and readiness in communication capabilities need testing. The current system would greatly benefit from a simulation that helps the SETI scientific community test their instruments and pipeline using an agreed-upon external source that acts as a trigger. The project would also prepare the scientific community to optimise communication strategies to effectively analyse the signal, allowing more antennas to receive it within a shorter time frame. Reception of the signal by multiple antennas is crucial for SETI search, as this would allow scientists to rule out natural or human-made triggers, confirming that the signal is effectively generated by an extraterrestrial civilisation.

- Operational Readiness: To conduct a realistic, large-scale "first contact" exercise to stress-test communication and response protocols.

  There are general guidelines set by the IAA SETI Committee in the Post Detection Protocol (2025-26) that regulate the detection of a radio signal, to ensure that all participating entities operate in alignment with these guidelines. There is however, no international technological or communication protocol that coordinates all radio telescopes to facilitate or confirm a signal detection. The test will enable us to help produce an accurate and reliable Detection Protocol that can optimise reception and detection capabilities. The test will allow the scientific community to detect strengths and weaknesses of current procedures to establish a more effective reception pipeline, and to establish and/or verify internal facility protocols for handling "highly suspicious signals."

- Infrastructure & Technology: To expand the global telescope network and upgrade facilities with real-time processing capabilities.

  Currently, only a few radio telescopes conduct structured SETI search, partly because the field has been marginalised within the scientific community for decades and partly because of a lack of dedicated funding. As we don’t know how and at what frequency a possible extraterrestrial signal might be received, it is crucial to expand the network using cost-effective and high-quality equipment to enable more facilities to receive a potential signal (such as NVIDIA Holoscan). Deploying these state-of-the-art technical solutions in telescopes participating in this activity is very challenging, in part due to the heterogeneity of these telescopes, but we believe that trying to demonstrate these capabilities is a step in the right direction. The Allen Telescope Array has recently been conducting some proof-of-concept testing of using Holoscan for SETI signal processing, in partnership with NVIDIA, and we are interested in expanding these capabilities as part of the project.

**Conclusions**

A Sign in Space part 2 aims to establish some of the scientific scopes and methodologies employed in A Sign in Space, including rehearsing the communication and technological pipelines coordinating various radio telescopes around the world, in the detection of an extraterrestrial signal.  A Sign in Space part 2 is primarily conceived as a scientific and technological preparedness exercise, helping the SETI community optimally prepare for a real extraterrestrial signal detection event. By bringing together radio observatories, space-science institutions, SETI researchers and technological partners, the project can transform hypothetical post-detection scenarios into measurable tests of existing capabilities. The outcomes generated from these exercises can strengthen detection methodologies, improve international coordination, guide infrastructure development, and contribute to the establishment of a more robust global SETI detection pipeline. In doing so, the project represents a practical step towards ensuring that, should a compelling technosignature ever be detected, the scientific community is technically capable, organisationally prepared, and internationally coordinated to evaluate the evidence with the highest possible degree of rigour. A Sign in Space part 2 is expected to foster an ongoing, long-lasting collaboration between scientists based at Earth-based radio observatories and space scientists at ESA, paving the way for an innovative chapter in SETI technological and scientific development.